\documentclass[a4paper]{article}

\usepackage{INTERSPEECH2021}

\usepackage{amsmath,graphicx,multirow,tabularx,booktabs,array}
\usepackage{color,soul}
\usepackage{xcolor}
\usepackage{array}
\usepackage{bbm}

\def\x{{\mathbf x}}
\def\y{{\mathbf y}}
\def\z{{\mathbf z}}

\title{Reducing Streaming ASR Model Delay with Self Alignment}
\name{Jaeyoung Kim$^1$, Han Lu$^1$, Anshuman Tripathi$^1$, Qian Zhang$^1$,  Hasim Sak$^1$}

\address{
  $^1$Google Inc., USA}
\email{\{jaeykim, luha, anshumant, zhaqian,  hasim\}@google.com}

\begin{document}

\maketitle
\begin{abstract}
Reducing prediction delay for streaming end-to-end ASR models with minimal performance regression is a challenging problem. Constrained alignment is a well-known existing approach that penalizes predicted word boundaries using external low-latency acoustic models. On the contrary, recently proposed FastEmit is a sequence-level delay regularization scheme encouraging vocabulary tokens over blanks without any reference alignments. Although all these schemes are successful in reducing delay, ASR word error rate (WER) often severely degrades after applying these delay constraining schemes. In this paper, we propose a novel delay constraining method, named self alignment. Self alignment does not require external alignment models. Instead, it utilizes Viterbi forced-alignments from the trained model to find the lower latency alignment direction. From LibriSpeech evaluation, self alignment outperformed existing schemes: $25\%$ and $56\%$ less delay compared to FastEmit and constrained alignment at the similar word error rate. For Voice Search evaluation, $12\%$ and $25\%$ delay reductions were achieved compared to FastEmit and constrained alignment with more than $2\%$ WER improvements.
\end{abstract}

\section{Introduction}

End-to-end ASR models~\cite{graves2006connectionist,graves:12,zhang2020transformer} have successfully expanded their presence by showing performance as competitive as conventional hybrid models~\cite{sainath2015convolutional}. In particular, recurrent neural network transducer (RNN-T) architecture~\cite{graves:12} gained the most popularity among them due to its natural streaming capability as well as superior performance. Transformer-Transducer~\cite{zhang2020transformer, yeh2019transformertransducer,tripathi2020transformer} further improved RNN-T architecture by replacing LSTMs with Transformer layers~\cite{vaswani2017attention}. 
However, streaming end-to-end models which optimize sequence likelihoods without any delay constraints suffer from high delay between the audio input and the predicted text because models learn to improve their prediction by using more future context.

Recently, there have been several approaches to reduce prediction delay. The constrained alignment~\cite{sak2015acoustic, sainath2020emitting, inaguma2020minimum} penalizes word boundaries based on audio alignment information from an external alignment model by masking out alignment paths exceeding the predetermined threshold delay. Inaguma~\textit{et al.}~\cite{inaguma2020minimum} proposed frame-wise CE regularization by constraining encoder output using a conventional hybrid model. 
FastEmit~\cite{yu2020fastemit}, unlike teacher-assisted training schemes, does not need external alignments. It is a sequence-level delay regularization scheme which encourages vocabulary tokens over blank one across the entire alignment paths in the RNN-T decoding graph.  

Although previous delay improving schemes can reduce latency of streaming end-to-end models, there are several issues on them. Teacher-assisted schemes such as constrained alignment heavily depend on the performance of external models. Since performance regression always happens when the model latency decreases, high-precision external alignment models would be needed to minimize WER degradation, which can further complicate model training steps. On the contrary, FastEmit blindly reduces delay by choosing the most efficient direction in the RNN-T decoding graph. However, its direction might not be optimal for all audio input due to lack of alignment information,  which can degrade delay-WER trade-offs. 

In this paper, we propose a novel delay constraining method, self alignment. Self alignment does not require external alignment models similar to FastEmit. However, the main difference is that it does not blindly optimize delay but utilizes Viterbi forced-alignments from the trained model to find the better low latency direction. For example, self alignment always finds the path one frame left to the Viterbi forced alignment and optimize it with the main objective. Therefore, self alignment keeps pushing the main alignment path to its left direction for each training step.

The proposed self alignment scheme has  advantages over existing schemes. First, training complexity for self alignment is much lower than teacher-assisted schemes since it does not need external alignment models.
Second, self alignment minimally affects ASR training by only constraining the most probable alignment path. On the contrary, other schemes affect many alignment paths by masking out them or changing weights on their label transition probabilities. Since delay constraining regularization terms always conflict with the main ASR loss, minimal intervention on the main loss would be important to optimizing delay and performance trade-offs. Self alignment only regularizes single path by pushing it to its left direction.
Third, self alignment provides better low latency direction than FastEmit.
From LibriSpeech evaluation, self alignment provides $25\%$ and $56\%$ less delay compared to FastEmit and constrained alignment when WER is fixed at around $4\%$. For large data evaluation, self alignment still provides better delay-WER trade-offs on Voice Search data: $12\%$ and $25\%$ delay reductions compared to FastEmit and constrained alignment, meanwhile maintaining better WER performance.  

\section{Transformer Transducers}
\begin{figure}[h]
\centering
\includegraphics[width=0.4\columnwidth]{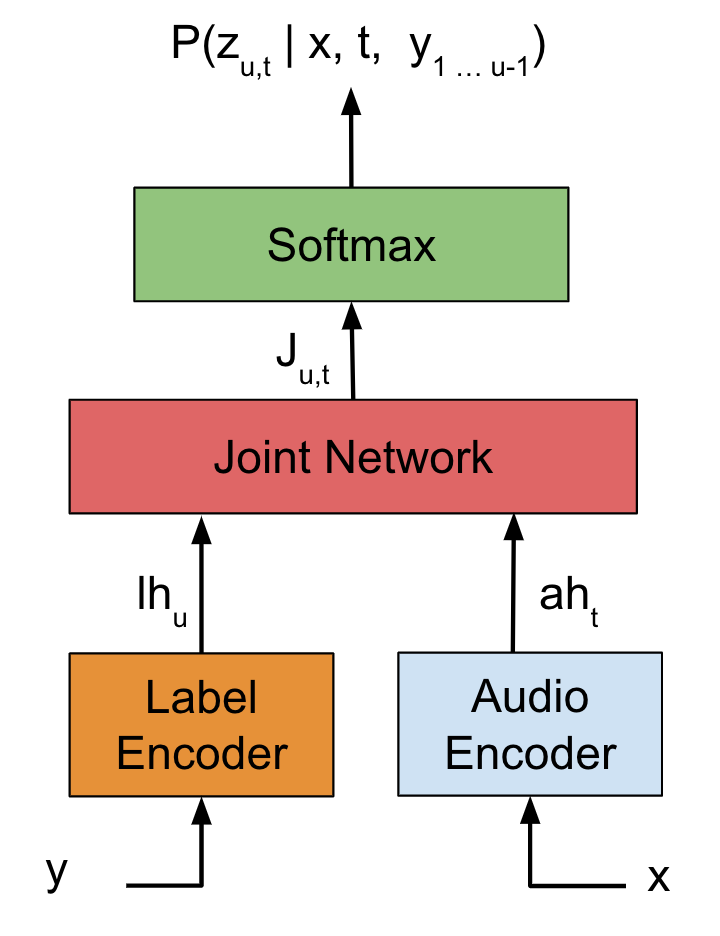}
\caption{Transformer Transducer architecture.}
\label{fig:rnnt_arch}
\end{figure}

Transformer-Transducer (T-T) was introduced in~\cite{zhang2020transformer, yeh2019transformertransducer,tripathi2020transformer} by replacing RNN-based audio and label encoders with Transformer models in RNN-T architecture~\cite{graves:12}. Figure~\ref{fig:rnnt_arch} depicts T-T architecture. For each time step, a T-T model provides alignment distribution $P(\z|\x)$ based on the past labels and audio signals which can be factorized as follows:
\begin{align}
    \textrm{Pr}(\z|\x) = \prod_i \textrm{Pr}(z_i|\x, t_i, \y_{1:u^{i}})
\end{align} 
where $\x$ is audio input, $\y$ is a ground-truth label sequence, $\z$ is an alignment belonging to $\y$, $\y_{1:u^i}$ is a partial label sequence processed from an alignment $\z_{1:(i-1)}$ by filtering blank symbols. Then the log conditional probability of $\y$ given audio input $\x$ is to sum all the alignment distributions corresponding to $\y$ as follows:
\begin{align}
    \log \textrm{Pr}(\y|\x) = \log \sum_{ K(\z) \in \y} \textrm{Pr}(\z|\x)
    \label{eq:prob_y_x}
\end{align} 
where the mapping $K$ removes blank symbols in $\z$.

The log total alignment probability in Eq.~\ref{eq:prob_y_x} is a target loss function which
can be efficiently computed using forward-backward algorithm as follows: 
\begin{align}
    \textrm{Pr}(\y|\x) = \alpha(T,U)
    \label{eq:forward1}
\end{align} 

\begin{align}
    \label{eq:forward2}
    \alpha (t,u) = \alpha(t-1, u-1) \textrm{Pr}(\phi | t-1, u) + \\ \nonumber
    \alpha(t,u-1)\textrm{Pr}(y_{u} | t, u-1)
\end{align} 
where $\textrm{Pr}(\phi | t-1, u)$ and $\textrm{Pr}(y_{u} | t, u-1)$ are blank and label probabilities and $T$ and $U$ are audio and label sequence lengths.

\section{Alignment Delay}
\begin{figure}[h]
\centering
\includegraphics[width=0.9\columnwidth]{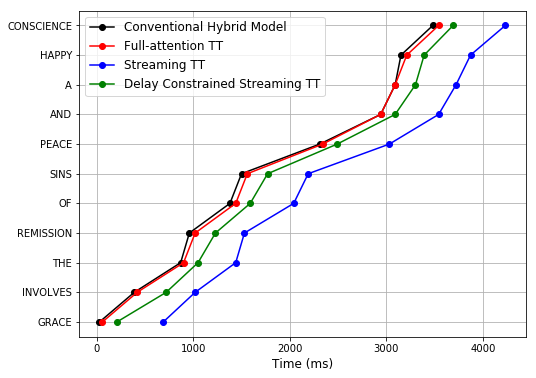}
\caption{Evaluation of alignment delay}
\label{fig:align_delay}
\end{figure}

\begin{table}[t]
  \centering
  \setlength{\tabcolsep}{2.0pt} 
  \begin{tabularx}{0.48\textwidth}{l r r r r}
    \toprule
    Streaming T-T & 
    Mean Delay & 
   RMS Delay & 
   TestClean & 
   TestOther \\
     \midrule
     baseline & 610ms &615ms & \textbf{3.4\%} & \textbf{9.5\%} \\
     delay const. & \textbf{172ms} & \textbf{175ms} & 4.4\% & 11.9\%\\
     \midrule
  \end{tabularx}
  \caption{Delay and WER comparison between streaming models with and without applying constrained alignment: Both average and RMS delays are evaluated on 100 Libri speech test clean samples.}
  \label{tab:compare_1}
\end{table}

In this section, we compare alignment delay between conventional acoustic model with cross entropy training based on Viterbi forced-alignments~\cite{sainath2015convolutional} and T-T models with the total alignment probability as in Eq.~\ref{eq:prob_y_x}.
Alignment delay is delay between input audio frames and streamed decoded labels.
Since label alignments can be gradually enhanced by iteratively training aligning models with realigned labels, a conventional model can learn accurate alignments after multiple iterations. In this section, alignments from a conventional model are assumed to be correct and used as ground-truth.

Figure~\ref{fig:align_delay} shows Viterbi forced-alignments for different T-T models. The x-axis is audio frame time in milliseconds and the y-axis is a label sequence.
A full-attention T-T model can access future frames when it computes self-attentions at Transformer layers. Its alignment path in Figure~\ref{fig:align_delay} almost coincides with the one from the conventional model. However, streaming T-T whose self-attention only depends on past frames showed excessive delay, more than 600 ms compared to full-attention T-T.
Since the total alignment probability in Eq.~\ref{eq:prob_y_x} includes all the label alignments without any delay constraint, a T-T model can arbitrarily increase its prediction delay in order to improve model accuracy by accessing to future context. 

Basically, the alignment delay is a critical issue for streaming ASR models. There are several existing delay constraining schemes for end-to-end training. The green alignment path at Figure~\ref{fig:align_delay} came from one of them, constrained alignment described at Section~\ref{subsec:constrained}.
After applying constrained alignment, the overall delay is significantly reduced from 610 ms to 172 ms at Table~\ref{tab:compare_1}. 
However, the model accuracy for the delay constrained model severely degraded due to the reduced future context. The mean and RMS delay definitions are explained at Section~\ref{subsec:setup}.
The delay constrained model reduced average delay around $440$ ms but test clean WER degraded close to $30\%$. Therefore, in order to properly evaluate alignment schemes, both delay and WER should be considered together.

\section{Related Works}
\label{sec:related}
\subsection{Constrained Alignment}
\label{subsec:constrained}
Constrained alignment training was first introduced for CTC models in~\cite{sak2015acoustic} and later extended to RNN-T and monotonic chunkwise attention (MoChA) models~\cite{sainath2020emitting,inaguma2020minimum}. It constrains the total alignment probability loss in Eq.~\ref{eq:prob_y_x} by completely  masking alignment paths with delay exceeding predetermined threshold. The delay is measured from the ground-truth alignment which can be supplied from external alignment models. There is also an extension to soft constrained alignment~\cite{sainath2020emitting}, where paths outside of the threshold delay are scaled by weights inversely proportional to the distance from the ground-truth alignments. 

\begin{figure}[h]
\centering
\includegraphics[width=0.9\columnwidth]{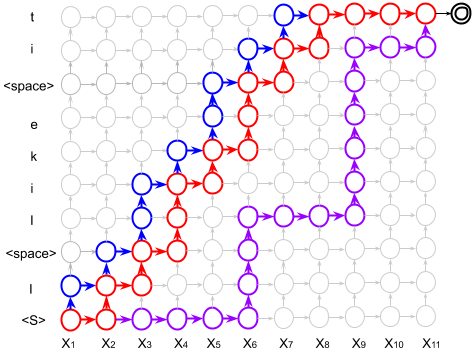}
\caption{T-T decoding graph for a label sequence, "I like it": The red alignment path is a reference alignment derived from either an external alignment model or delay-constrained model itself.}
\label{fig:viterbi_graph}
\end{figure}

Figure~\ref{fig:viterbi_graph} shows an example of a decoding graph for a label sequence 'I like it'. 
The threshold delay is enforced only to a word boundary token, a space symbol. 
The red alignment path at Figure~\ref{fig:viterbi_graph} is a reference alignment and the purple alignment is the rightmost allowed path when word boundary threshold is set to be 2. The constrained alignment for word boundary tokens can be formulated in the forward algorithm as follows: 
\begin{align}
    \label{eq:const}
    \alpha (t,u) = & \alpha(t-1, u-1) \textrm{Pr}(\phi | t-1, u) + \\ \nonumber
                   & \alpha(t,u-1)\textrm{Pr}(y_{u} | t, u-1) \mathbbm{1}{\left(t<T_{u} + \sigma\right)}
\end{align} 
where $T_{u}$ is ground-truth alignment time for $u^{\textrm{th}}$ token,  $\sigma$ is threshold and  $\mathbbm{1}$ is an indicator function. Eq.~\ref{eq:const} is only applied when $u^{\textrm{th}}$ label is a word boundary token. Otherwise, the original forward algorithm at Eq.~\ref{eq:forward2} is used.

\subsection{FastEmit}
FastEmit is a newly proposed delay-constrained scheme~\cite{yu2020fastemit}. It does not require external alignments, which is a main advantage over constrained alignment or encoder distillation. It directly manipulates forward-backward sequence probabilities by boosting the probability of the next vocabulary token and discouraging the blank transition at all $(u,t)$ positions. For example, FastEmit pushes alignment paths to the upper left direction in Figure~\ref{fig:viterbi_graph}. FastEmit training can be implemented by simple gradient boosting technique as in Eq.~11 and Eq.~12 from~\cite{yu2020fastemit}. 

\section{Proposed Scheme}

We propose a new delay improving scheme, named self alignment. The self alignment scheme does not need external alignment models. Instead, Viterbi forced-alignments from the trained model itself are used to find the delay improving direction, which is one frame left to the forced-alignment path. For example, in Figure~\ref{fig:align_delay}, the red alignment path is the forced-alignment from the trained model and the blue alignment is one frame left to it. 
For each training batch, the self alignment scheme is to encourage the left alignment paths, which constantly pushes the model's forced-alignments to the left direction. The training loss can be formulated as follows:
\begin{align}
    \mathcal{L}_{\textrm{total}} = -\log \textrm{Pr}(\y|\x) -  \lambda \sum_{u} \log \textrm{Pr} \left( y_u | t_u,u \right) 
    \label{eq:selfalign}
\end{align} 
where $\lambda$ is a weighting factor for the left-alignment likelihoods,  $t_u$ is a frame index for the left alignment at the $u^{\textrm{th}}$ token.

\begin{figure}[h]
\centering
\includegraphics[width=0.85\columnwidth]{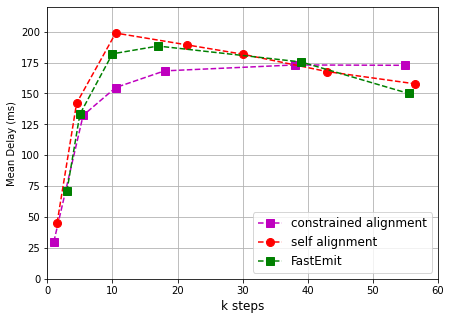}
\caption{Comparison of delay training curves}
\label{fig:training_curve}
\end{figure}

The proposed self alignment scheme has advantages over the existing schemes seen at Section~\ref{sec:related}. First, like FastEmit, it does not need external alignment models. Constrained alignment needs external teacher models for reference alignments and, therefore, their performance heavily depends on the quality of teacher models. Second, self alignment can push alignments for the training model to the better direction than FastEmit. FastEmit provides good delay-WER trade-off without increasing memory and computational complexity. However, FastEmit blindly pushes alignments to the left upper direction in the decoding graph, which is clearly not optimal direction for all the audio input.
Self alignment utilizes its own forced-alignments to guess the next alignment direction. 


Figure~\ref{fig:training_curve} compared mean delay for each training steps. For constrained alignment, delay regularization is strictly applied to all training steps because mean delay is monotonically increasing. Delay regularization is usually conflict with main loss and the strict regularization for all training steps would make training hard for constrained alignment. On the other hand, FastEmit and self alignment showed delay constraint is not dominant for initial training steps. Mean delays initially increase for minimizing main RNN-T loss but when main loss is close to convergence, delay starts to decrease where delay regularization is getting dominant. This trend especially makes sense for self alignment. Initially, training is focused on the main loss but when Viterbi forced alignments are getting to make sense, training is shifting to reduce the delay of the Viterbi paths.


\section{Experiments and Results}

\subsection{Setup}
\label{subsec:setup}
We evaluate delay improving schemes on two separate datasets: LibriSpeech~\cite{panayotov2015librispeech} and Google's Voice Search. LibriSpeech corpus is a read speech data based on audio books and consists of 1000 hours of training and test sets. For larger data evaluation, we use Google Voice Search corpus with more than 30k hours. The test set contains 14k Voice Search utterances with length less than 5.5 seconds. The training and test sets are all anonymized and hand-transcribed. 
The input audio is processed as 128 dimension logmel energy features and stacked with 4 frames. Before fed as acoustic features, stacked logmels are subsampled by a factor of 3, resulting in 30ms features. For robust training, specaugment~\cite{park2019specaugment} is applied to acoustic features.

We use two delay metrics to compare different schemes: 1) mean alignment delay and 2) root mean square (RMS) delay. The mean alignment delay is defined as mean word time difference between ground-truth and predicted alignments:   
\begin{align}
    D_{\textrm{mean}} = \frac{1}{\sum_{k=1}^N |\y_k|}\sum_{k=1}^{N} \sum_{i=1}^{|\y_k|} (\hat{t}^k_i - t^k_i)
\end{align} 
where $t^k_i$ is $i^{\textrm{th}}$ word time from a reference model, $\hat{t}^k_i$ is $i^{\textrm{th}}$ predicted word time, $|\y_k|$ is $k^{\textrm{th}}$ utterance length and $N$ is the number of utterances.
One issue for the mean alignment delay is that it cannot reflect delay variance in the metric. RMS delay can measure delay variance and defined as follows:
\begin{align}
    D_{\textrm{rms}} = \sqrt{\frac{1}{\sum_{k=1}^N |\y_k|}\sum_{k=1}^{N} \sum_{i=1}^{|\y_k|} (\hat{t}^k_i - t^k_i)^2}
\end{align} 

\subsection{Main Results on Libri Speech}
\label{subsec:libri}
\begin{figure}[t]
    \centering
    \includegraphics[width=0.9\columnwidth]{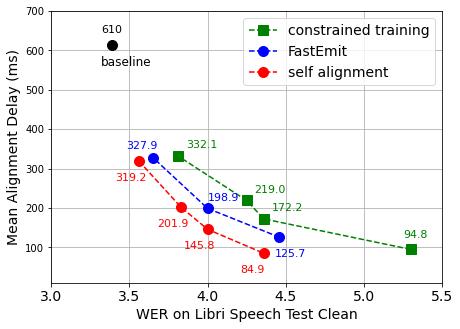}
    \caption{Delay-WER trade-off comparison on Libri speech test clean}
    \label{fig:delay_wer_tradeoff}
    \vspace{-11pt} 
\end{figure}

\begin{table}[t]
  \centering
  \setlength{\tabcolsep}{2.0pt} 
  \begin{tabularx}{0.5\textwidth}{l r r r r}
    \toprule
    Model & 
    Mean Delay & 
   RMS Delay & 
   TestClean & 
   TestOther \\
     \midrule
     \textbf{A} Streaming T-T & 610ms & 615ms & 3.4\% & 9.5\% \\
     \midrule
    \textbf{A} + ConstAlign & 328ms & 330ms &4.0\% & 11.1\%\\
    \textbf{A} + FastEmit & 195ms &215ms &4.0\% & \textbf{10.4\%} \\
    \textbf{A} + SelfAlign & \textbf{145ms}& \textbf{164ms} & 4.0\% & 10.7\% \\
     \midrule
  \end{tabularx}
  \caption{Delay and WER evaluation on LibriSpeech at around $4\%$ test clean WER: Both mean and RMS delays were evaluated on 100 LibriSpeech test clean samples.}
  \label{tab:libri_wer_delay}
\end{table}

The streaming T-T audio encoder for LibriSpeech consists of 15 Transformer layers with 100 left context frames for each layer. The streaming T-T model is trained with an output delay of 4 frames. The label encoder has 2 Transformer layers with 20 left context symbols. 

Figure~\ref{fig:delay_wer_tradeoff} compares delay-WER trade-offs on LibriSpeech test clean data. Delay-WER trade-offs can be obtained by sweeping delay tuning hyper-parameters. For example, constrained alignment can adjust the model delay by tuning $\sigma$ at Eq.~\ref{eq:const}. For other schemes, $\lambda$ serves as a tuning hyper-parameter. In Figure~\ref{fig:delay_wer_tradeoff}, self alignment outperformed all other schemes from low to high latency region. For high latency region, FastEmit and self alignment showed similar delay-WER trade-offs. Their gap gradually increases as the model moves into the low latency region. 
Constrained alignment showed poor delay-WER trade-offs, which is a surprising result considering it utilized external alignment models. Full-attention T-T was used as an alignment model which can have higher variance compared to a conventional model because a full-attention T-T was not directly trained with alignment labels. The second reason is that delay constraint is imposed for all training steps for constrained alignment which could make training harder than other schemes. Although not used in this evaluation, soft constrained alignment~\cite{sainath2020emitting} could make training more easier because it does not completely mask out outside of the delay threshold.

Table~\ref{tab:libri_wer_delay} compared mean and RMS delay on LibriSpeech at around $4\%$ WER. Self alignment showed $25\%$ and $56\%$ less mean delay compared to FastEmit and constrained alignment, respectively.   

\subsection{Evaluation on Voice Search Corpus}
\label{subsec:vs}
\begin{figure}[t]
    \centering
    \includegraphics[width=0.9\columnwidth]{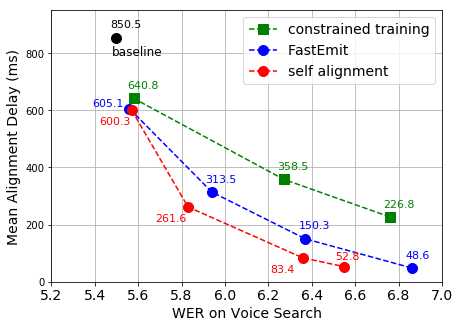}
    \caption{Delay-WER trade-off comparison on Voice Search.}
    \label{fig:vs_delay_wer_tradeoff}
    \vspace{-11pt} 
\end{figure}

\begin{table}[t]
  \centering
  \begin{tabularx}{0.46\textwidth}{l r r r r}
    \toprule
    Model & 
    Mean Delay & 
  RMS Delay & 
  VS WER \\
     \midrule
     \textbf{A} Streaming T-T & 856ms & 858ms & 5.5\%  \\
     \midrule
    \textbf{A} + ConstAlign & 358ms & 360ms &6.3\% \\
    \textbf{A} + FastEmit & 307ms &331ms &5.9\%  \\
    \textbf{A} + SelfAlign & \textbf{269ms}& \textbf{278ms} & \textbf{5.8\%}  \\
     \midrule
  \end{tabularx}
  \caption{Delay and WER evaluation on VS: Both mean and RMS delays were evaluated on 100 LibriSpeech test clean samples.}
  \label{tab:compare_constrained_or_not}
\end{table}

In this section, we evaluate self alignment on the larger audio corpus, Voice Search (VS) data. The audio encoder for Voice Search has 18 Transformer layers with 32 left context frames. The label encoder has the same architecture as for Libri Speech.
Figure~\ref{fig:vs_delay_wer_tradeoff} compares Delay-WER curves for constrained training, FastEmit and self alignment. Similar to the LibriSpeech result, self alignment outperformed other schemes for all latency region. For high latency region, FastEmit and self alignment showed less than $0.1\%$ WER loss with more than 250ms delay reduction. As moving into the lower latency region, their WER gap gradually increases. Constrained training still showed poor delay-WER trade-off compared to self alignment and FastEmit.
Table~\ref{tab:compare_constrained_or_not} compared different schemes with mean delays around $300$ms. Self alignment showed $12\%$ and $2\%$ relative improvements on delay and WER over FastEmit. Compared to the constrained alignment, self alignment presented $25\%$ and $7\%$ relative gains on delay and WER, respectively.

\section{Conclusions}
\label{sec:conclusions}
In this paper, we proposed a novel delay constraining method, self alignment. Self alignment uses Viterbi forced-alignments from the trained model to find the lower latency alignment direction. Since it does not use external alignment models, training steps are much simpler than the constrained alignment scheme. Moreover, Viterbi forced-alignments can provide the better delay improving direction than FastEmit. The experimental result showed that the proposed self alignment significantly outperformed existing approaches on delay-WER trade-offs for both LibriSpeech and Voice Search datasets.

\bibliographystyle{IEEEtran}

\bibliography{template}


\end{document}